\documentclass[aps,english,10pt,twocolumn]{revtex4}
\usepackage{amsmath}
\usepackage{amsthm}
\usepackage{amssymb}
\usepackage{amsfonts}
\usepackage{bbm}
\usepackage{epsfig}
\usepackage{times}
\usepackage{babel}
\usepackage{color}
\usepackage{hyperref}
\usepackage{framed}
\usepackage{changes}
\usepackage{float}
\usepackage{array}

\begin{document}

\title{Out-of-time-order correlators in the one-dimensional XY model}
\author{Jiahui Bao$^{2}$}
\email{baojh5@mail2.sysu.edu.cn}
\author{Cheng-Yong Zhang$^{1}$}
\email{zhangcy@jnu.edu.cn}
\affiliation{$^{1}$Department of Physics and Siyuan Laboratory, Jinan University, Guangzhou 510632, China}
\affiliation{$^{2}$School of Physics and Astronomy, Sun Yat-sen University, 2 Daxue Rd., Zhuhai 519082, China}


\begin{abstract}
Out-of-time-order correlators (OTOC) are considered to be a promising tool to characterize chaos in quantum systems. In this paper we study OTOC in XY model. With the presence of anisotropic parameter $\gamma$ and external magnetic field $\lambda$ in the Hamiltonian, we mainly focus on their influences on OTOC in thermodynamical limit. We find that the butterfly speed $v_B$ is dependent of these two parameters, and the recent conjectured universal form which characterizes the wavefront of chaos spreading are proved to be positive with varying $v_B$ in different phases of XY model. Moreover, we also study the behaviors of OTOC with fixed location, and we find that the early-time part fully agrees with the results derived from Hausdorff-Baker-Campbell expansion. The long-time part is studied either, while in the local case $C(t)$ decay as power law $t^{-1}$, $|F(t)|$ with nonlocal operators show quite interesting and nontrivial power law decay corresponding to different choices of operators and models. At last, we observe temperature dependence for OTOC with local operators at ($\gamma=0, \lambda=1$), and divergent behavior with low temperature for nonlocal operator case at late time.
\end{abstract}

\maketitle

\section{Introduction}
Chaos is an interesting phenomenon in quantum system. It's closely related to the nature of quantum mechanics and black holes, making it attractive in several physics communities like  condensed matter physics, quantum information and high energy physics.
Recently, a conjecture was proposed to establish a bound on strong quantum chaos \cite{Maldacena:2015waa}, also named quantum butterfly effect \cite{Shenker:2013pqa}, who characterizes the behavior of chaos as $\sim e^{\lambda_Lt}$. The Lyapunov exponent $\lambda_L$ is unbounded for classical systems and bounded as $\lambda_L\leq2\pi k_BT/\hbar$ for quantum systems. Systems saturate this bound are called fast scramblers \cite{Sekino:2008he, Lashkari:2011yi}, like black holes. However, there also exist some models that don't show such exponential growth \cite{Huang:2016knw, Chen:2016qpx, Slagle:2016fnd, Fan:2016ean, PhysRevB.95.024202}, which are known as slow scramblers. These many-body quantum systems include rich information about the relation between thermalization and information scrambling, which may also relate to the study of hiding information behind black hole horizon. Therefore, it's important to understand their properties by the observations from both theoretical and experimental ways.

Quantum chaos can be characterized by Out-of-Time-Order Correlator (OTOC), which captures critical information differed from two-point correlation function. It can also be understood as the measurement of delocalization of spreading operators. Consider following quantity
\begin{equation}
C(l,t)=\frac{1}{2}\langle[W(l,t),V(0)]^{\dagger}[W(l,t),V(0)]\rangle\,,
\end{equation}
where $\langle ... \rangle\equiv \langle e^{-\beta H} ...\rangle/ \langle e^{-\beta H}\rangle$ denotes thermal average at temperature $T=1/\beta$ and $W(l,t)\equiv e^{iHt}W(l)e^{-iHt}$.
Assuming operators $W$ and $V$ are both unitary and Hermitian, we can rewrite it as $C(l,t)=1-Re[F(l,t)]$,
where $F(l,t)=\langle W(l,t)V(0)W(l,t)V(0)\rangle$ is OTOC for its special time ordering. It attracts a lot of attention not only because its richness in theoretical physics, but also because its feasibility in experiments \cite{Swingle:2016var, Zhu:2016uws, kaufman2016quantum, garttner2017m, landsman2018verified, lukin2018probing}.

The behavior of OTOC has several interesting aspects. First, the early-time behavior is usually characterized by Hausdorff-Baker-Campbell (HBC) formula.
Besides, when it comes to the area around the wavefront, there is a conjectured universal form to describe the ballistic broadening of OTOC \cite{Xu:2018xfz, Khemani:2018sdn}
\begin{equation}
C(l,t)\sim {\rm exp}(-c\frac{(l-v_Bt)^{1+p}}{t^p})\,,\label{eq:wavefront}
\end{equation}
where $c$ is constant and $v_B$ is spreading velocity of butterfly effect, which indicates the speed that wavefront propagates. It's decided by setting velocity-dependent Lyapunov exponents $\lambda_L(v_B)=0$, thus the region of chaos spreading can be divided into two parts $v>v_B$ and $v<v_B$, and only classical or semi-classical systems show exponential growth behavior inside the wavefront. Moreover, the Eq.\eqref{eq:wavefront} usually only holds outside the wavefront. Besides, $p$ is a coefficient related to models. For example, $p=1$ for random circuit model \cite{Nahum:2017yvy, vonKeyserlingk:2017dyr, Khemani:2017nda}, $p=1/2$ for non-interacting translation-invariant model \cite{Xu:2018xfz, Lin:2018luj, Jian:2018ies}, $p=0$ for Sachdev-Ye-Kitaev (SYK) model \cite{Sachdev:1992fk} and chains of coupled SYK dots at large $N$ \cite{Gu:2016oyy}.
At last, the long time part also have interesting behaviors, which may reveal important information about how operators saturate the chaos bound.

Thus, in order to understand the behavior of chaos spreading and verify the conjectured form, it's promising to calculate OTOC in different systems, including integrable or chaotic ones. Recently, some work have been done to analyze the OTOC  in conformal field theories \cite{Roberts:2014isa, Stanford:2015owe, Roberts:2014ifa, Chowdhury:2017jzb, Patel:2017vfp}, quantum phase transition \cite{Shen:2016htm, sun2018out}, Luttinger liquids \cite{Dora:2017tbd}, and also some lattice integrable models like quantum Ising chain \cite{Lin:2018tce}, hard-core boson model \cite{Lin:2018luj}, quadratic fermions \cite{Byju:2018eyb}, random field XX spin chain \cite{Riddell:2018}  and symmetric Kitaev chain\cite{mcginley2018slow}. Scrambling was observed in critical point of Ising spin chain for nonlocal operators, and weak chaos was also witnessed in some models. It's well known that both quantum Ising model and XX model can be seen as special cases of XY model \cite{Lieb:1961fr}, who possesses an extra parameter $\gamma$ that denotes the difference of component in $x$ and $y$ direction for two nearest neighbours coupling. As this anisotropy property is common in real physical systems, and XY model itself has many nontrivial quantum phase transitions and properties \cite{sachdev2011quantum}, it's interesting to study OTOC in XY model, especially the behaviors of operator growth and information scrambling.

In this paper we focus on the evolution of OTOC in XY model, including its butterfly velocity and wavefront universal form Eq.\eqref{eq:wavefront}. We find that the butterfly velocity is dependent of $\gamma$ and $\lambda$, and with this varying velocity the universal form holds perfectly for all OTOC and phases in XY model. In addition to this, we also study the early time and long time behavior of OTOC, while the former is characterized by HBC formula, the later shows quite interesting and unusual power law behaviors. Interesting temperature dependence is also observed in particular cases.

This paper is organized as follows. We will introduce XY model in Section \ref{II}, including its quantum phase transition and procedures to diagonalize the Hamiltonian. In Section \ref{III}, the calculation method of OTOC will be outlined, and then we will show exactly how it evolves with time and space, in order to extract the information behind the calculation. Then in Section \ref{IV} we will briefly discuss these results and conclude.

\section{XY model} \label{II}

\begin{figure}[htbp]
	\centering
	\includegraphics[width=0.25\textwidth]{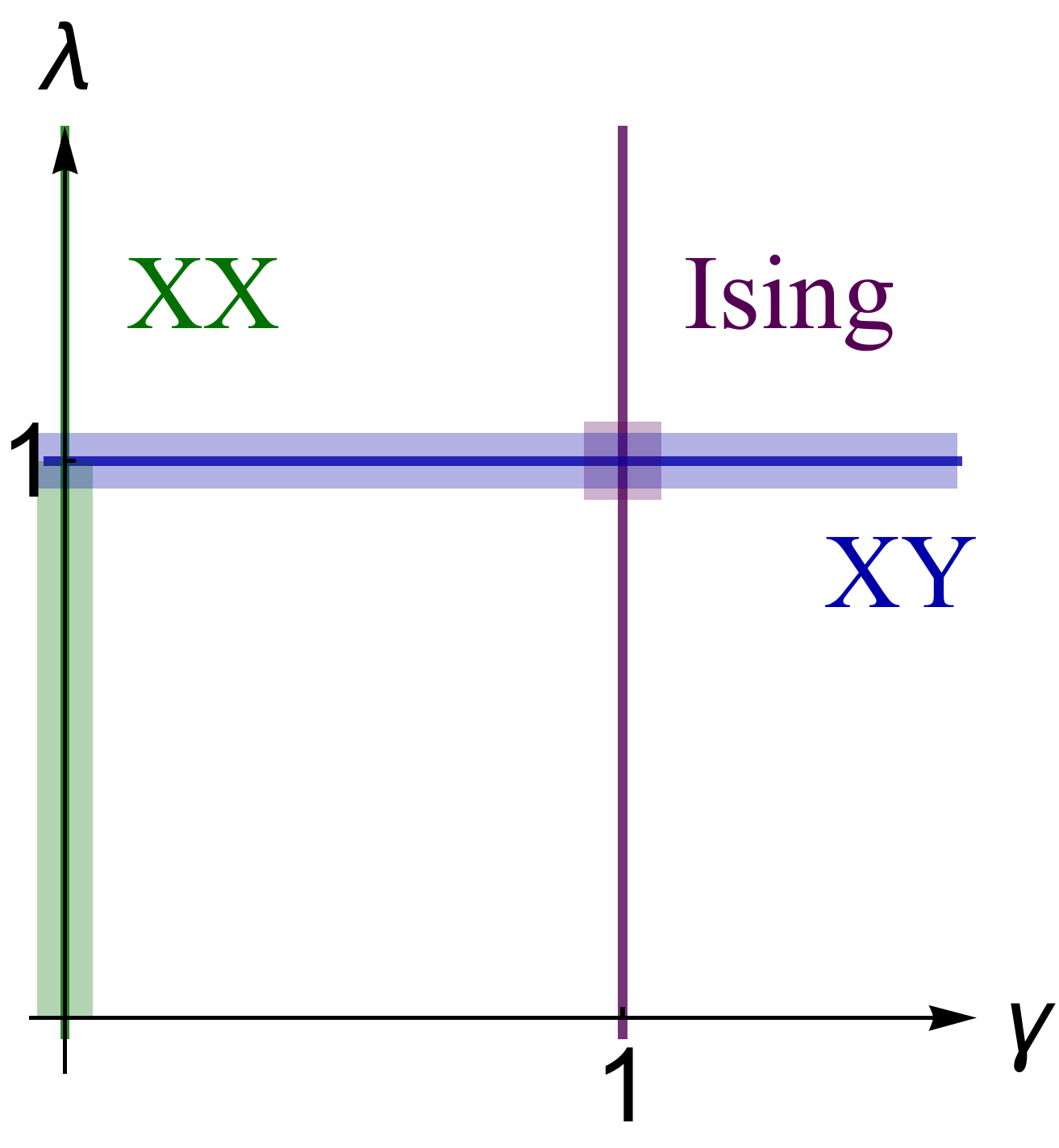}
	\caption{Critical regions of XY model from \cite{Latorre:2003kg}, colored shadow area is the corresponding critical region of each model. We only plot one quadrant because the existence of symmetry $\gamma\rightarrow-\gamma$ and $\lambda\rightarrow-\lambda$.}\label{XYphase}
\end{figure}

XY model is one of the simplest nontrivial integrable model, it has rich phase diagram and potential ability to study new effects. The Hamiltonian of it is
\begin{equation}
H=-\frac{J}{2}\sum_{j=0}^{N-1} \Big{[}\frac{1+\gamma}{2}\sigma_j^x\sigma_{j+1}^x +\frac{1-\gamma}{2}\sigma_j^y \sigma_{j+1}^y+\lambda\sigma_j^z \Big{]}\,,
\end{equation}
where $\gamma$ is anisotropy coefficient, describes the difference of interactive strength in the $x$ and $y$ components, and $\lambda$ describes magnetic field along $z$ direction. These two parameters decide the phases and properties of this model, when $\gamma=0$ it becomes isotropic XY model (also called XX model), and when $\gamma=1$ it recovers quantum Ising chain. The relationship between them is shown in Fig.\ref{XYphase}, and the shadow areas are the corresponding critical regions of different models.

The critical regions at $\lambda=1$ (blue line) and $\gamma=0, 0\leq\lambda\leq1$ (green line) are conformal invariant, correspond to conformal charge $c=1/2$ CFT and $c=1$ CFT, respectively. Thus, XY chain has two quantum phase transitions located at these two lines, where the spectrum becomes gapless. And the line located at $\lambda=1$ is a transition from doubly degenerate state ($\lambda<1$) to single ground state ($\lambda>1$).  However, the point (0, 1) is not conformal since the dynamical critical exponent of it is equal to 2 \cite{franchini2017introduction}. Overall, its interesting properties can be revealed further by analyzing its OTOC.

In order to calculate the OTOC of XY chain, we should diagonalize its Hamiltonian using Jordan-Wigner transformation and Bogoliubov transformation first. We set $J=1$ for general energy scale, and rewrite the Pauli matrices by spin operators $\sigma_j^x=a_j^{\dagger}+a_j$, $\sigma_j^y=(a_j^{\dagger}-a_j)/i$, $\sigma_j^z=2a_j^{\dagger}a_j-1$ together with Jordan$-$Wigner transformation $a_j=\big{[}{\rm exp}\big{(}i\pi \sum_{l=1}^{j-1}c_l^{\dagger}c_l\big{)}\big{]}c_j$, the Hamiltonian will become
\begin{equation}
\begin{split}
&H=-\frac{1}{2}\sum_{j=0}^{N-1} \Big{[}(c_j^{\dagger}c_{j+1}-c_jc_{j+1}^{\dagger})+\gamma(c_j^{\dagger}c_{j+1}^{\dagger}-c_jc_{j+1})\\
&+\lambda(2c_j^{\dagger}c_j-1) \Big{]}+\frac{\mu}{2}\Big{(}c_N^\dagger c_0+c^\dagger_0 c_N+\gamma c_N^\dagger c_0^\dagger+\gamma c_0 c_N\Big{)}\,,
\end{split}
\end{equation}
where $\mu=\prod_{j=1}^N \sigma_j^z$ is the parity operator. In order to deal with the boundary term, the Hamiltonian can be separated as
\begin{equation}
\begin{split}
H=&\frac{1+\mu}{2} H^++\frac{1-\mu}{2}H^-\,,\\
H^{\pm}=&-\frac{1}{2}\sum_{j=0}^{N-1} \Big{[}(c_j^{\dagger}c_{j+1}-c_jc_{j+1}^{\dagger})\\
&+\gamma(c_j^{\dagger}c_{j+1}^{\dagger}-c_jc_{j+1})+\lambda(2c_j^{\dagger}c_j-1) \Big{]}\,,
\end{split}
\end{equation}
since the even/odd parity of their number is conserved.

Therefore, with this form, we are able to use appropriate Fourier transform $c_k=\frac{e^{-i\pi/4}}{\sqrt{N}}\sum_{j=0}^{N-1}e^{-ijk}c_j$ and Bogoliubov transformation $\gamma_k={\rm cos}\theta_k c_k -{\rm sin}\theta_k c_{-k}^{\dagger}$ to complete the diagonalization as
\begin{equation}
\begin{split}
H^\pm&=\sum_{k^\pm}\epsilon_{k^\pm}(\gamma_{k^\pm}^{\dagger}\gamma_{k^\pm}-\frac{1}{2})\,,\\
k^\pm&=\frac{2\pi[n+\frac{1\pm1}{4}]}{N}\,, n=0,1,...N-1\,,
\end{split}
\end{equation}
where $\epsilon_k=[({\rm cos}k-\lambda)^2+\gamma^2{\rm sin^2}k]^{1/2}$ is dispersion of the elementary excitations and the Bogoliubov angle $\theta_k$ satisfies
$
{\rm tan}(2\theta_k)=\frac{\gamma\,{\rm sin}k}{\lambda-{\rm cos}k}.
$

Since we need to calculate the thermal average of operators, we need to know how to use this diagonalized Hamiltonian to do it. Actually, it has been studied in \cite{Lin:2018tce} that in thermodynamical limit $N\rightarrow\infty$, we have $\langle O\rangle=\langle O\rangle_+=\langle O\rangle_-$ for OTOC with either local or nonlocal operators in Ising model, here the subscript denotes the choice of $k^\pm$ corresponding to even/odd chain length $N$. We have checked this conclusion holds in XY model, and in this paper we will use $k^+$ and even $N$ for consistency.

\section{out-of-time-order correlator} \label{III}

With the diagonalized Hamiltonian we are able to calculate OTOC of XY model now. Choosing different combinations of $(\gamma,\lambda)$, and using Pauli matrices to replace operators $W$ and $V$, we need to calculate following term
\begin{equation}
F_{\mu\nu}(l,t)=\langle\sigma^{\mu}_l(t)\sigma^{\nu}_0\sigma^{\mu}_l(t)\sigma^{\nu}_0\rangle\,,
\end{equation}
where $\mu,\nu=x,y,z$. The Pauli matrices can be expressed by fermionic operators in Majorana representation $A_j=c_j^{\dagger}+c_j$ and $B_j=c_j^{\dagger}-c_j$:
\begin{equation}
\begin{split}
&\sigma_j^x=(\prod_{j'<j}A_{j'}B_{j'})A_j,\\
&\sigma_j^y=-i(\prod_{j'<j}A_{j'}B_{j'})B_j,\\
&\sigma_j^z=-A_jB_j.\label{Marep}
\end{split}
\end{equation}

Now all the OTOC can be expressed as thermal average of Majorana fermions sequences. For instance, $F_{zz}(l,t)=\langle A_l(t)B_l(t)A_0B_0A_l(t)B_l(t)A_0B_0\rangle$, and in thermodynamical limit, it can be computed using Wick's theorem, which turns the calculation of long sequence into combination of two-point correlation functions. We will use Pfaffian method here to do the calculation numerically with similar steps in \cite{Lin:2018tce}.

The Pfaffian method \cite{lieb1968theorem, bravyi2012disorder} can be expressed as
\begin{equation}
F(l,t)=\pm{\rm Pf}(\Phi)=\pm\sqrt{{\rm Det}(\Phi)}\,,\label{Pfa}
\end{equation}
where the matrix $\Phi$ is skew-symmetric, i.e. $\Phi_{ii}=0$ and $\Phi_{ij}=-\Phi_{ji}$. This form will be modified if we use ``double trick'' to deal with the calculation, and the sign of $F(l,t)$ is not definitely positive then. But it can still be decided by requiring the ``continuity'' of OTOC, which we will elaborate later. The matrix $\Phi$ is constructed in terms of Majorana correlation functions, $\Phi_{ij}=\langle X_iX_j \rangle$, where $X_i$ is $i-$th element inside thermal average function $\langle X_1X_2... \rangle$.

Therefore, the basic correlation functions are $\langle A_m(t)A_n\rangle$, $\langle A_m(t)B_n\rangle$, $\langle B_m(t)A_n\rangle$ and $\langle B_m(t)B_n\rangle$, which can be derived if we know the exact diagonalized form of Hamiltonian, and their expressions are shown in Appendix ~\ref{app:Majorana}.

\subsection{OTOC with local operators}

OTOC characterises the chaos spreading and information scrambling, in other words, the delocalization of operators. And since the behavior of many-body localized quantum chaos can be revealed by the local operators, study of them becomes quite interesting.

\begin{figure}[!htbp]
	\centering
	\includegraphics[width=0.45\textwidth]{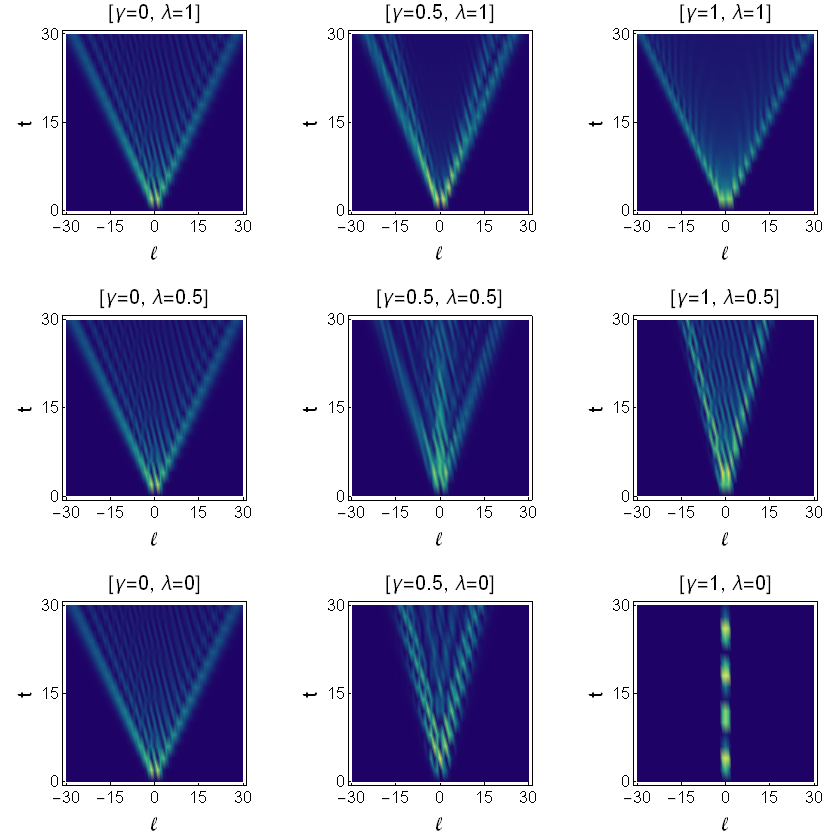}
	\caption{General evolution of $C_{zz}$ with different choices of parameters in XY model. We set system size $N=500$, inverse temperature $\beta=0$. And the coordinates are space ranges from -30 to 30, time ranges from 0 to 30 respectively. Lighter color means bigger value of $C_{zz}$, corresponding to stronger delocalization. These pictures show clearly how the spreading of operator is bounded by ``cone structure''.}\label{fig:localSP}
\end{figure}

For XY model, the OTOC with local operators is $C_{zz}$, as shown in Eq.\ref{Marep}, operator $\sigma_j^z=-A_jB_j$ is local because it's consisted of fermions only located at site $j$, and $\sigma_j^x$ and $\sigma_j^y$ are nonlocal for their connection with all sites of fermions before site $j$. Following Eq.\ref{Marep}) and Eq.\ref{Pfa}, we have
\begin{equation}
\begin{split}
C_{zz}(l,t)&=1-{\rm Re}[\langle (A_l(t)B_l(t)A_0B_0)^2\rangle]\\
&=1-{\rm Re}\sqrt{{\rm Det}(\Phi_{zz})}\,,
\end{split}
\end{equation}
with Pfaffian trick, we can compute this quantity numerically.
To study how two key parameters $(\gamma, \lambda)$ affect the general evolution of OTOC and the spreading velocity of the butterfly effect, we illustrate the results using typical choices of their values in Fig.\ref{fig:localSP}. Here we choose system size $N=500$, $\beta=0$, so the temperature is infinite. Moreover, lighter color denotes stronger $C(t)$, thus it shows how OTOC spreads. Several interesting properties can be observed from these results.

First, the cone structure, which indicates the bound of butterfly effect, is observed except the model with $\gamma=1, \lambda=0$. It corresponds to quantum Ising chain without external magnetic field, and the Hamiltonian is $H=-J/2\sum \sigma_j^x\sigma_{j+1}^x$. Therefore, $C(t)$ of this model is always zero at locations expect 0 and $\pm1$, which means operator doesn't spread in this case. Actually, It can be understood by considering the HBC formula, the expansion of operators with time
\begin{equation}
\begin{split}
W(t)&=\sum_{n=0}^{\infty}\frac{(it)^n}{n!}L_n(W)\\
&=W+it[H,W]+\frac{(it)^2}{2!}[H,[H,W]]+...\,.
\end{split}
\end{equation}
For this case, the commutator $[W(l,t),V(0)]=[\sigma_l^z(t),\sigma_0^z]$ will vanish for all sites except $l=0, \pm1$, so $C(l,t)\equiv\frac{1}{2}\langle|[W(l,t),V(0)]|^2\rangle$ will also vanish, too.

Second, the spin chain with $(\gamma=1$, $\lambda=1)$ and $(\gamma=0$, $0<\lambda<1)$ always satisfy that the butterfly effect $v_B=1$, and in the later case OTOC actually are not influenced by magnetic field $\lambda$ if $\gamma=0$ and $\beta=0$. Nevertheless, the other positions have narrower cone structure, indicating that their speed of operator spreading is relatively slower. Thus, it seems that $v_B$ is actually dependent of $\gamma$ and $\lambda$, but not constant in all cases. So, we can conclude that the existence of anisotropy of a system will affect its speed of operator spreading, for weak magnetic field, it will slow the speed down. Furthermore, the external magnetic field will increase the velocity of spreading if $x$ and $y$ components are not equal.

\begin{figure}[!htbp]
	\centering
	\includegraphics[width=0.45\textwidth]{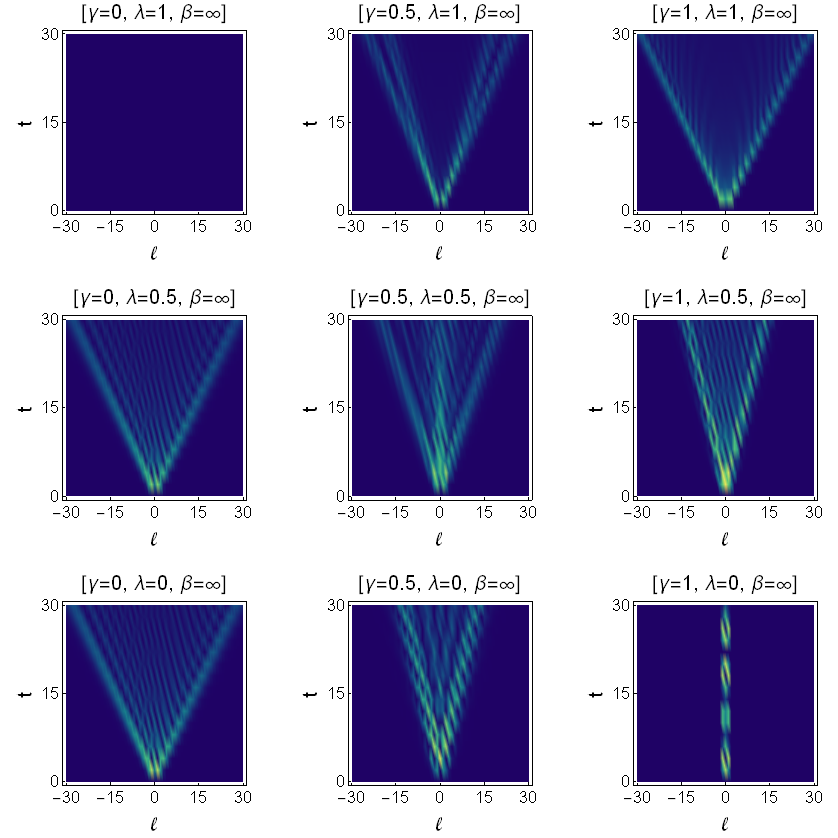}
	\caption{Evolution of $C_{zz}$ at zero temperature $T=1/\beta=0$ with other configurations same as Fig.\ref{fig:localSP}. It shows that only model of ($\gamma=0, \lambda=1$) has vanishing $C_{zz}$, all other models are not sensitive to temperature.}\label{fig:localinf}
\end{figure}

Third, the temperature has negligible effects on OTOC with local operators except the model with $(\gamma=0, \lambda=1)$, which as the temperature falls to zero, the OTOC will vanish. We show the evolution of $C_{zz}$ at $\beta=\infty$ in Fig.\ref{fig:localinf}. The reason of this phenomenon is not obvious, but we know that at this point, the model is not critical as mentioned in Section \ref{II}, so we suppose that this may be helpful to explain why this phenomenon exists.

\begin{figure}[!htbp]
	\centering
	\includegraphics[width=0.4\textwidth]{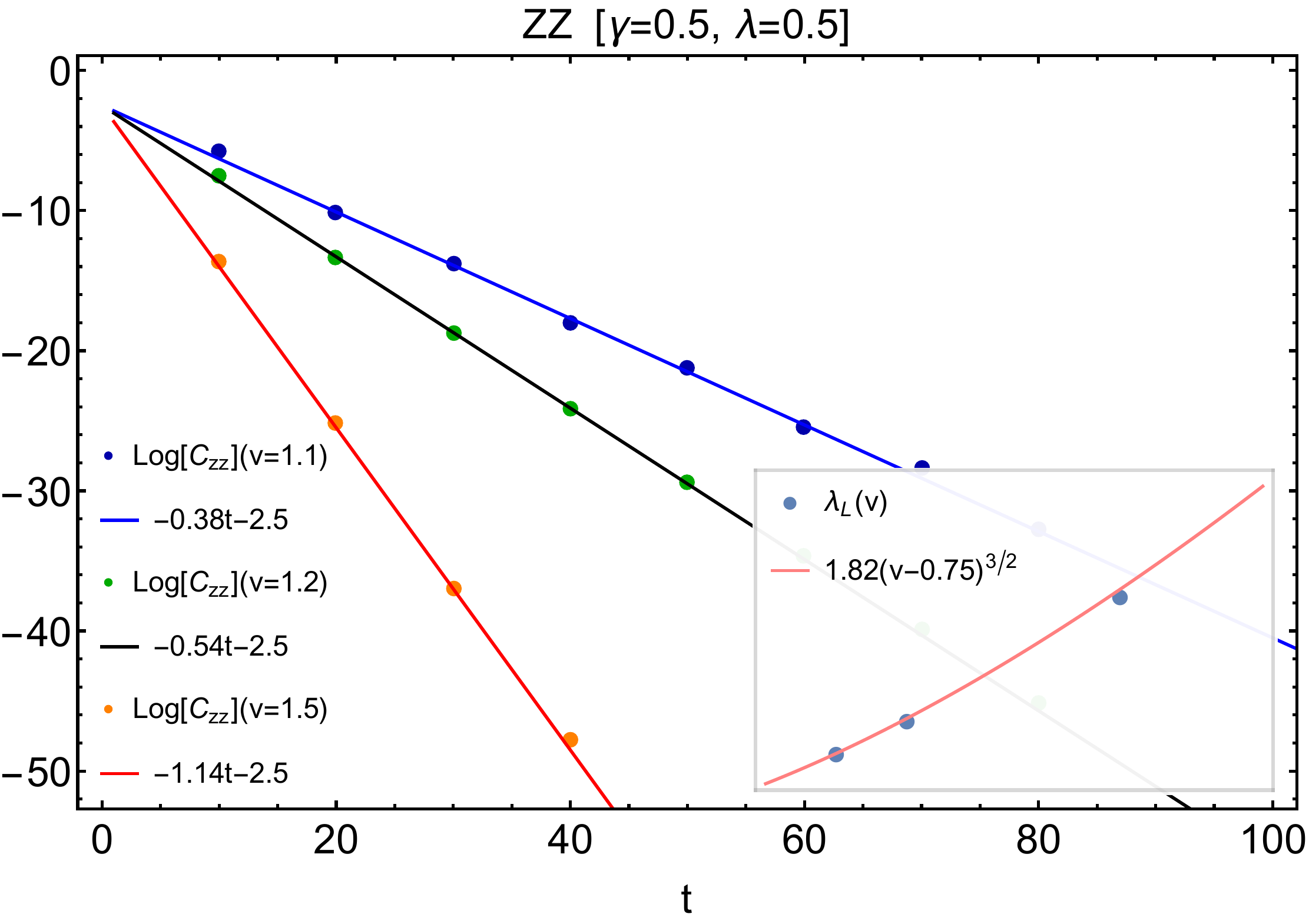}
    \includegraphics[width=0.4\textwidth]{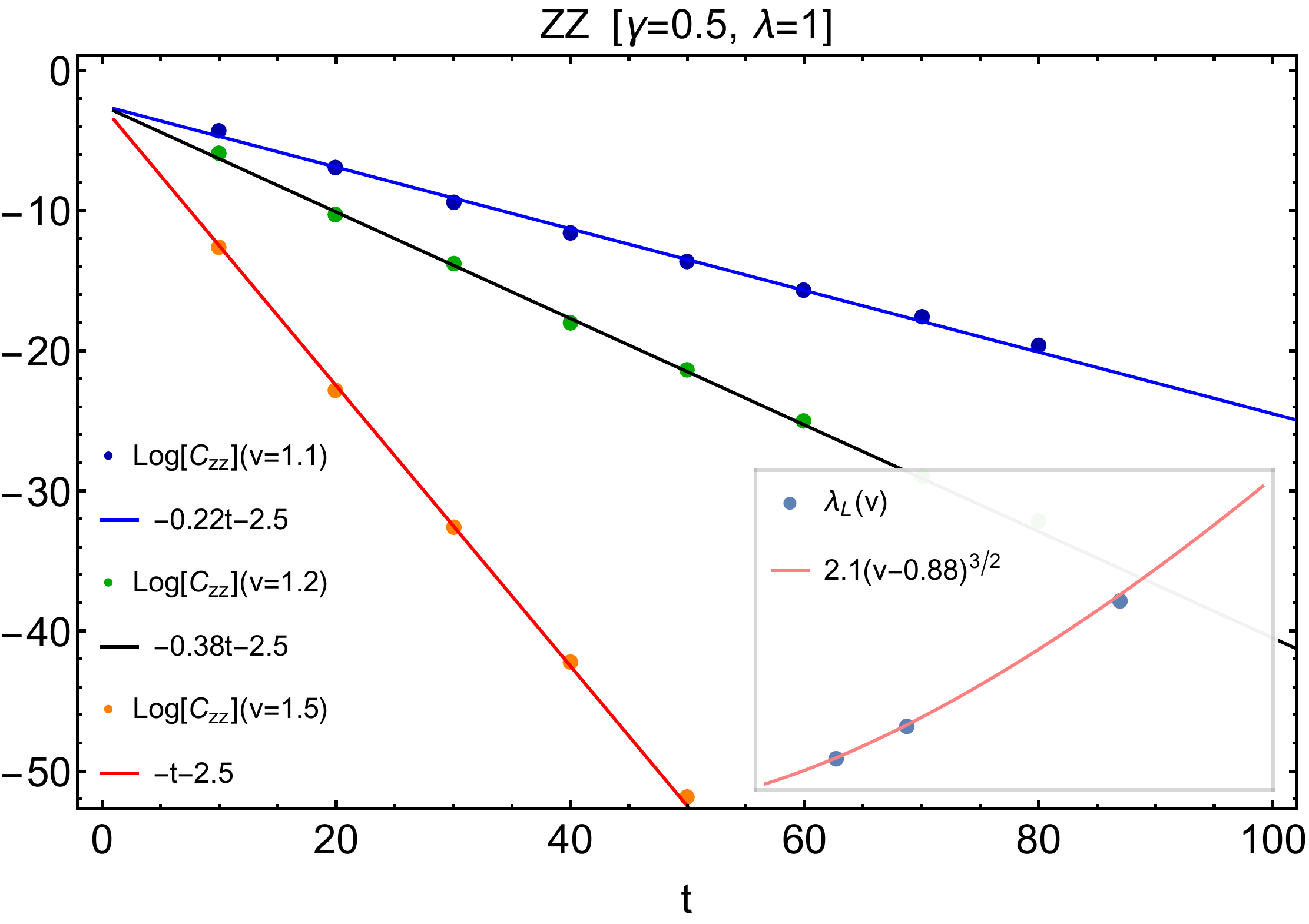}
    \includegraphics[width=0.4\textwidth]{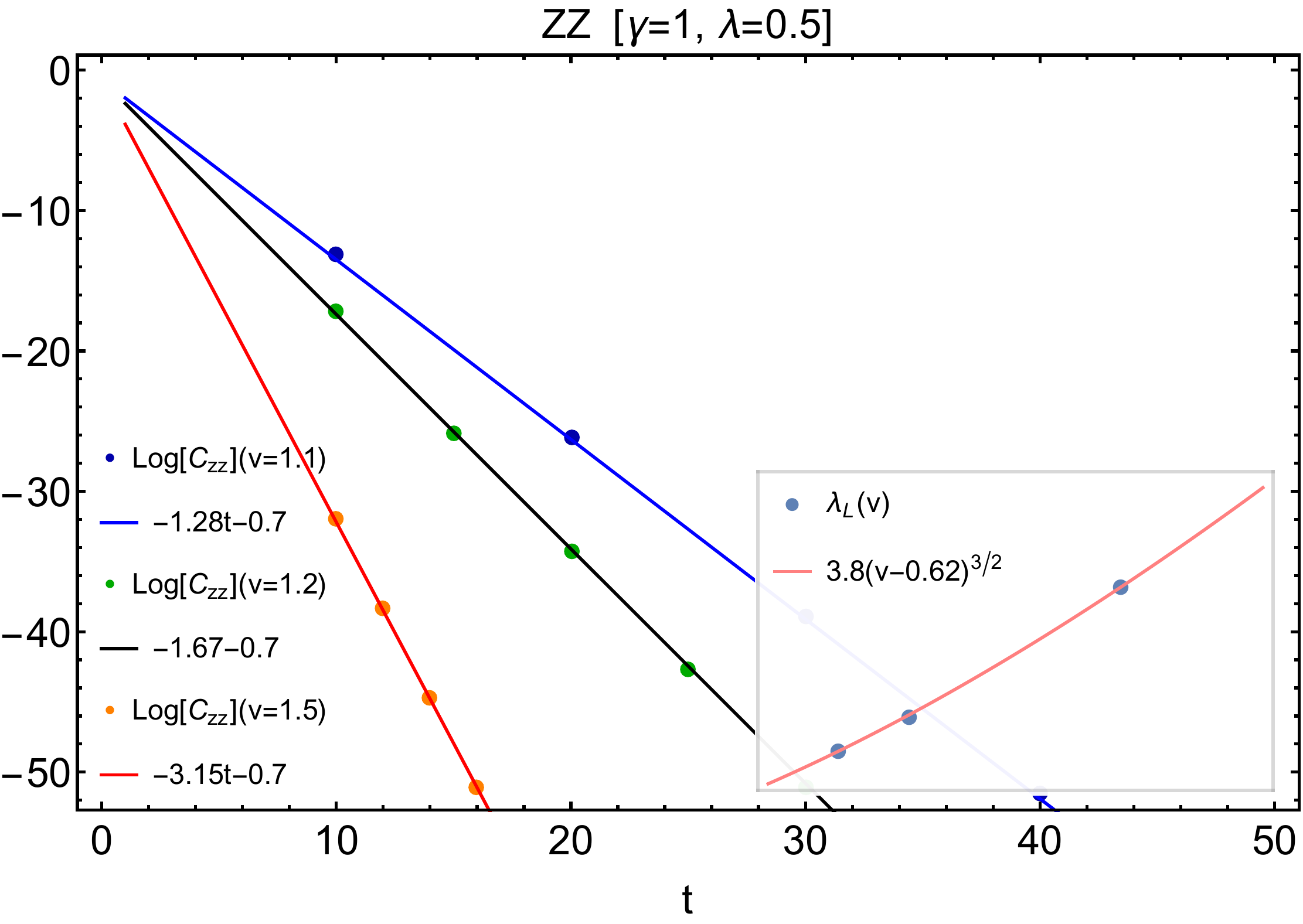}
	\caption{Fitting of the universal form and numerical data of $C_{zz}$. Here we pick three models with different $v_B$ to check whether the conjecture holds in XY model. The dots outside are numerical data by calculating the OTOC along velocity-fixed rays with $v=1.1, 1.2, 1.5$ respectively. And the solid lines are fitting forms of $-at+b$, $a$ is the velocity-dependent Lyapunov exponents $\lambda_L(v)$ we need to extract. The inset shows how this three sets of extracted data fitted with $\sim(v-v_B)^{1+p}$ as a function of $v$. We can see the numerical data fits quite well.}\label{fig:localwf}
\end{figure}

In addition to the general evolution of OTOC with local operators, there are more to be explored in XY chain. For instance, the universal form describing the wavefront of $C(t)$. This proposal suggests that around the wavefront of chaos spreading, where the velocity $v=l/t>v_B$, can be described by Eq. \eqref{eq:wavefront}.
In order to verify this conjecture with XY model, especially with the existence of $\gamma$ and $\lambda$, we need to calculate the wavefront of $C_{zz}$ along the fixed-velocity rays. Our results are shown in Fig.\ref{fig:localwf}, which describes the cases of $C_{zz}$ with three sets of parameters $(\gamma=0.5, \lambda=0.5)$, $(\gamma=0.5, \lambda=1)$ and $(\gamma=1, \lambda=0.5)$. We choose them because their velocities of spreading are different.

Here we posit that
\begin{equation}
C(t)=const.\times {\rm exp}(-\lambda_L(v)t)\,,
\end{equation}
to check its relation with Eq.\eqref{eq:wavefront}. If the universal form indeed holds in XY model, then we can have the relation
\begin{equation}
\begin{split}
{\rm Log}[C(t)]&=-\lambda_L(v)t+const.\,,\\
\lambda_L(v)&\sim(v-v_B)^{1+1/2}
\end{split}
\end{equation}
as p=1/2 for XY model.

In Fig.\ref{fig:localwf}, we use time $t$ and ${\rm Log}[C_{zz}]$ as coordinates, with fixed velocity $v=1.1, 1.2, 1.5$. Then we use form $-\lambda_L(v)t+const.$ to fit the numerical results. In the inset pictures the values of $\lambda_L(v)$ are extracted by fitting the numerical data and the power law relation are checked to find out whether this conjecture holds in XY model. For example, in the top picture, the coordinates of three points are (1.1, 0.38), (1.2, 0.54) and (1.5, 1.14), which fits function $\lambda_L(v)=1.82(v-0.75)^{3/2}$ quite well. And we can see from Fig.\ref{fig:localSP}, the butterfly velocity of model $(\gamma=0.5, \lambda=0.5)$ is indeed $v_B\approx0.75$. The other two figures also support this relation quite well, so we can conclude for OTOC with local operators, the universal form is supported by XY model. On the other hand, this result also support our former conclusion that $v_B$ depends on $\gamma$ and $\lambda$.

\begin{figure}[!htbp]
	\centering
	\includegraphics[width=0.48\textwidth]{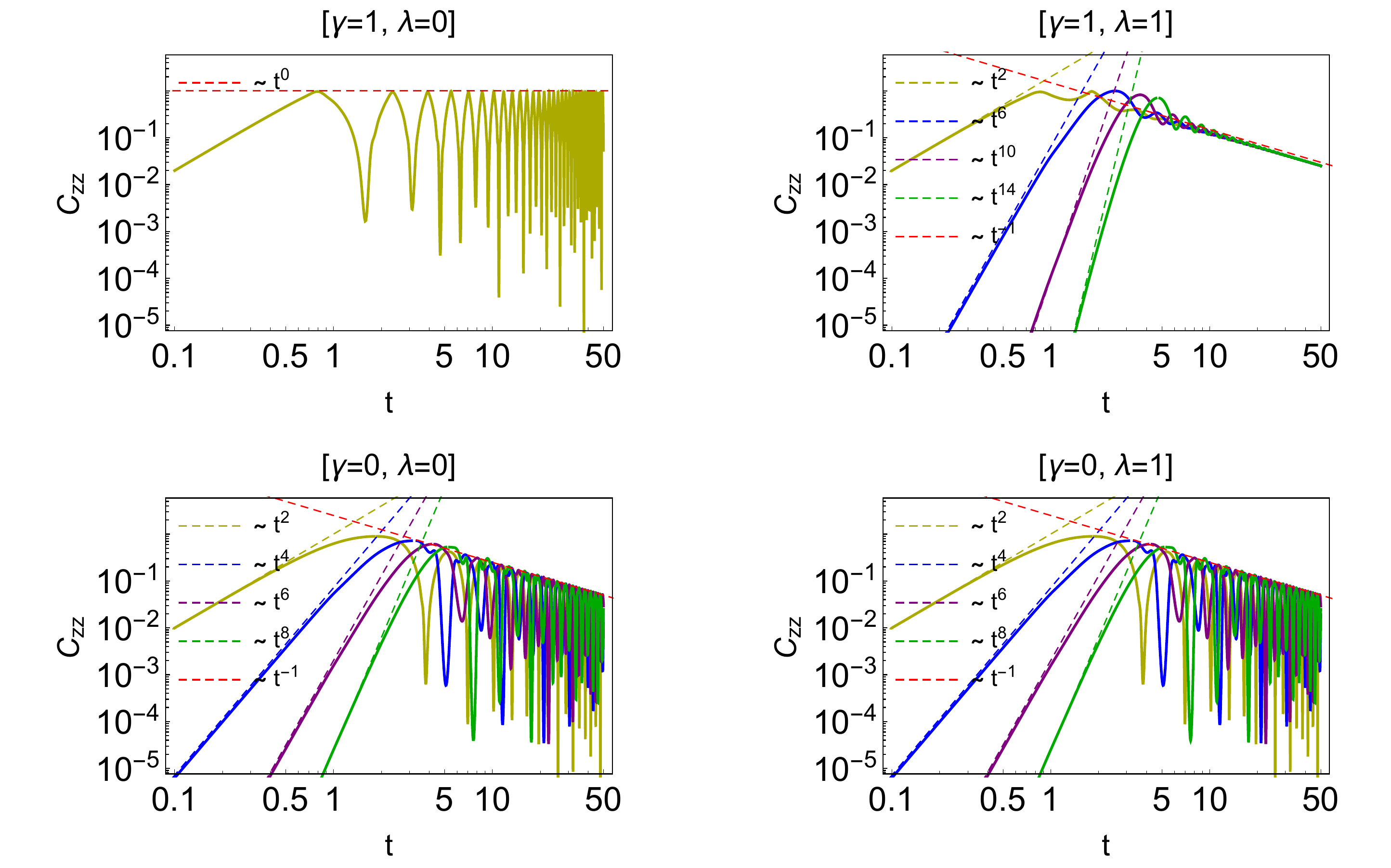}
	\caption{Time evolution of $C_{zz}$ with four typical models. Yellow, blue, purple, green lines correspond to fixed locations $l=1, 2, 3, 4$, respectively. $C_{zz}$ at ($\gamma=1$, $\lambda=0$) vanishes for $l>1$. The dashed lines are used for power law fitting, we can see clearly from the figures that these models show $t^{4l-2}$ and $t^{2l}$ power law growth at early time, and $t^{-1}$ decay at long time independent of site $l$.}\label{fig:ZZT}
\end{figure}

After study the wavefront part to verify the universal form, it's also important to study the time evolution behavior of $C_{zz}$ with fixed sites. It will tell us how the local operators behave exactly, in exponential or power law way. There are two meaningful parts to study, the early time and the long time. The results of models at four typical points are illustrated in Fig.\ref{fig:ZZT} with sites $l=1, 2, 3, 4$, and we can see clearly their early time behavior is vanishing at ($\gamma=1, \lambda=0$) for $l>1$, $t^{4l-2}$ for ($\gamma=1, \lambda=1$) and $t^{2l}$ for the rest. These behaviors can be understood by HBC formula, since time $t$ is small for early time, the description by HBC expansion is quite accurate. Notice the lowest order of $t$ which makes $C(t)$ nonzero is decided by the lowest order of $L_n(W)$ that satisfies $[L_n(W),V]\ne0$. To be specific, the Hamiltonian of XY model can be divided into six different kinds by choosing different values of $\gamma$ and $\lambda$, (1). $\sigma^x\sigma^x$, (2). $\sigma^y\sigma^y$, (3). $\sigma^x\sigma^x+\sigma^z$, (4). $\sigma^y\sigma^y+\sigma^z$, (5). $\sigma^x\sigma^x+\sigma^y\sigma^y$, (6). $\sigma^x\sigma^x+\sigma^y\sigma^y+\sigma^z$, For $zz$ OTOC, $\sigma^x\sigma^x$ and $\sigma^y\sigma^y$ have no difference, so there are four kinds of behaviors. We have known that type (1) and (2) have vanishing $C(t)$, and for type (3), the Ising spin chain, we have
\begin{equation}
\begin{split}
L_1(\sigma^z_0)&=[H,\sigma^z_0]\sim\sigma^y_0\sigma^x_1\,,\\
L_2(\sigma^z_0)&\sim[H,\sigma^y_0\sigma^x_1]\sim\sigma^y_0\sigma^y_1\,,\\
L_3(\sigma^z_0)&\sim[H,\sigma^y_0\sigma^y_1]\sim\sigma^y_0\sigma^z_1\sigma^x_2\,...\\
\end{split}
\end{equation}
So we can find the lowest order $[L_n(\sigma_0^z),\sigma_l^z]\ne0$ by checking above forms, and in this case, $[\sigma_0^z(t),\sigma_l]$ shows $t^{2l-1}$ power law growth, thus $C(t)=\frac{1}{2}\langle|[\sigma_0^z(t),\sigma_l]|^2\rangle$ has $t^{4l-2}$. Early time behaviors of other types can also be checked using similar method, and notice type (5) and type (6) show same $t^{2l}$ growth behavior because when x and y components exist at the same time, their behavior will dominate. Overall, we can conclude that the early time behaviors of $C_{zz}$ can be separated into three areas, $t^{4l-2}$ for ($\gamma=1,\lambda\ne0$), $t^{2l}$ for $\gamma\ne1$, and vanishing at ($\gamma=1, \lambda=0$) for $l>1$.

Now we can continue to study the late-time part of $C_{zz}$, as exhibited in Fig.\ref{fig:ZZT}, they all decay as $t^{-1}$, independent of $\gamma$, $\lambda$, site $l$ and temperature. This behavior can be understood by the stationary phase approximation of fermionic correlation functions, i.e. when $t\rightarrow \infty$, we have \cite{Lin:2018tce}
\begin{equation}
C_{zz}(l,t)\sim(1-\langle A_0B_0\rangle^2)\frac{2}{\pi|\epsilon^{''}_\pi|t}\,,
\end{equation}
where $\epsilon_\pi^{''}$ is the second derivative of $\epsilon_k$ with $k=\pi$.

\subsection{OTOC with nonlocal operators}

There are five kinds of OTOC with nonlocal operators in XY model, i.e. $F_{xx}$, $F_{yy}$, $F_{xy}$, $F_{xz}$ and $F_{yz}$.
However, since the operators $\sigma^x_j$ and $\sigma^y_j$ change the fermion parity, their Heisenberg evolution can't simply obtained from $A_j(t)$ and $B_j(t)$. Following the method used in \cite{Lin:2018tce, McCoy:1971zz}, we will use the ``double trick'' to deal with the calculation of OTOC. This method need to construct a new form, so the OTOC can be calculated by Wick's theorem directly.
Consider quantity
\begin{equation}
\Gamma_{\mu\nu}(j,t)\equiv\langle (\sigma_{\frac{N}{2}}^\mu(t)\sigma_{N-j}^\mu(t)\sigma_0^\nu\sigma_{\frac{N}{2}-j}^\nu)^2  \rangle\,,
\end{equation}
for large enough $N$, invoking Lieb-Robinson bound and cluster property \cite{McCoy:1971zz}, we can get
\begin{equation}
\begin{split}
\Gamma_{\mu\nu}(j,t)&\approx\langle (\sigma_{\frac{N}{2}}^\mu(t)\sigma_{\frac{N}{2}-j}^\nu)^2 \rangle \langle (\sigma_{N-j}^\mu(t)\sigma_{0}^\nu)^2 \rangle\\
&=F_{\mu\nu}(j,t)F_{\mu\nu}(-j,t)=F_{\mu\nu}^2(j,t)\,.\label{eq:Gamma}
\end{split}
\end{equation}
here $F_{\mu\nu}(j,t)=F_{\mu\nu}(-j,t)$ because of mirror symmetry.

Taking $xx$ OTOC as an example, we need to calculate quantity $\Gamma_{xx}(j,t)$, which can be expressed as
\begin{equation}
\begin{split}
\Gamma_{xx}(j,t)=&\langle \bigg{[}\Big{(}\prod_{j'=\frac{N}{2}}^{N-j-1} B_{j'}(t)A_{j'+1}(t)\Big{)}\\
&\times\Big{(}\prod_{j'=0}^{\frac{N}{2}-j-1} B_{j'}A_{j'+1}\Big{)}\bigg{]}^2  \rangle\,.
\end{split}
\end{equation}
Then we can use Pfaffian method to calculate it. First we need construct a matrix $\Phi_{xx}$ of dimension $4(N-2j) \times 4(N-2j)$, then compute $F_{xx}(j,t)$ as
\begin{equation}
\begin{split}
F_{xx}(j,t)=\pm \sqrt{|{\rm Pf}(\Phi_{xx})|}=\pm [{\rm Det}(\Phi_{xx})]^{\frac{1}{4}}\,.
\end{split}
\end{equation}
Since the quantity is doubled, we don't know the sign of it directly, but it can be recovered by requiring the ``continuity'' of $F_{xx}(j,t)$. More specifically, there is a critical rule for all the points on site $j$ and time $t$: on the premise of turning least directions, choose closest distance. With this rule we can check how the curve is finally organized with all the points from calculation. And we should notice that when ${j>vt}$, $F_{xx}\rightarrow1$ for the existence of light cone. OTOC with other operators can also be calculated in the same way. Following Eq.\ref{Marep} and Eq.\ref{eq:Gamma}, we have
\begin{equation}
\begin{split}
\Gamma_{xy}(j,t)=&\langle \bigg{[}\Big{(}\prod_{j'=\frac{N}{2}}^{N-j-1} B_{j'}(t)A_{j'+1}(t)\Big{)}\\
&\times\Big{(}\prod_{j'=0}^{\frac{N}{2}-j-1} A_{j'}B_{j'+1}\Big{)}\bigg{]}^2  \rangle\,,
\end{split}
\end{equation}
\begin{equation}
\begin{split}
\Gamma_{xz}(j,t)=&\langle \bigg{[}\Big{(}\prod_{j'=\frac{N}{2}}^{N-j-1} B_{j'}(t)A_{j'+1}(t)\Big{)}\\
&\times A_0A_{N/2-l}B_0B_{N/2-l}\bigg{]}^2  \rangle\,,
\end{split}
\end{equation}
$\Gamma_{yy}$ and $\Gamma_{yz}$ can also be constructed with this method.

\begin{figure}[!htbp]
	\centering
	\includegraphics[width=0.45\textwidth]{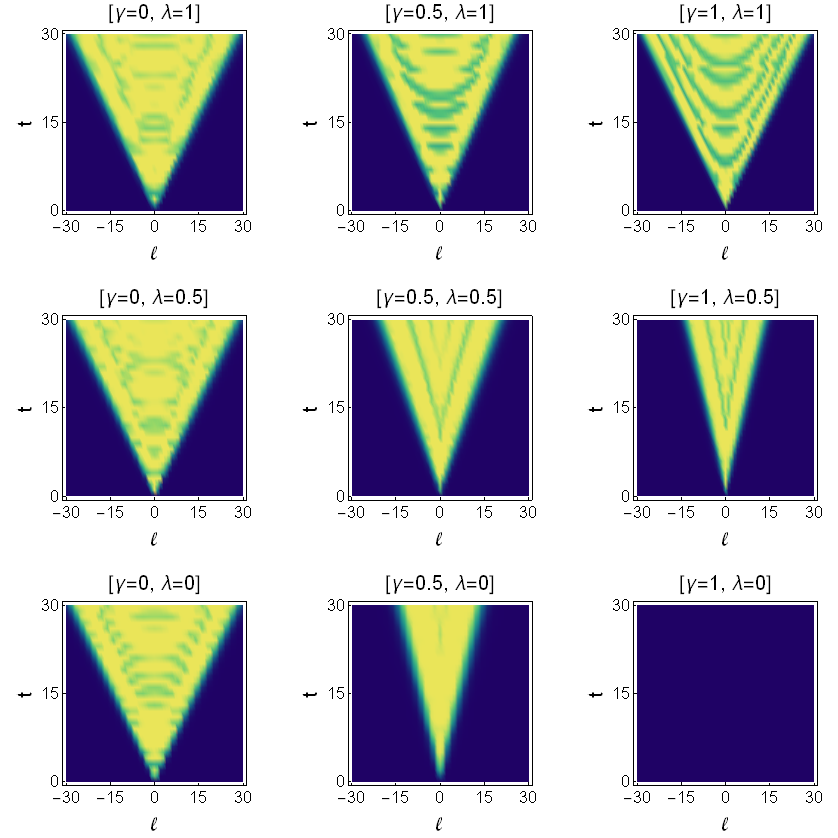}
	\caption{General evolution of $C_{xx}$ with different choices of parameters in XY model. We set system size $N=100$, inverse temperature $\beta=0$. And the coordinates are space ranges from -30 to 30, time ranges from 0 to 30 respectively. Lighter color means bigger value of $C_{xx}$, corresponding to stronger delocalization. These pictures show similar light cone bound for different models, but there are also some differences compared with the local case inside the light cone: it's relatively more scrambled instead of tranquility.}\label{fig:xxsp}
\end{figure}

Then we can compute similar quantities like before. The general behavior of $C_{xx}$ is illustrated in Fig.\ref{fig:xxsp} with system size $N=100$ and $\beta=0$. And other $C(t)$ with different operators show similar behaviors, so we don't show them here. Notice unlike the vanishing $C(t)$ in Fig.\ref{fig:localinf} at zero temperature with ($\gamma=0, \lambda=1$), it will not vanish for nonlocal case. Moreover, from these figures we can see that the butterfly velocity is the same as that in local case, which means the butterfly velocity depends only on the model but not the operators in OTOC function. Moreover, scrambling is observed for all sites inside the ``light cone''. And now when $\gamma=0, \beta=0$, $\lambda$ has tiny effect on OTOC. The scrambling observation of OTOC with nonlocal operators shows their main differences compared with local ones, and it can be easily understood, since nonlocal operators have nonlocal information about operators, which lead to delocalization once they spread inside the light cone.

\begin{figure}[!htbp]
	\centering
	\includegraphics[width=0.4\textwidth]{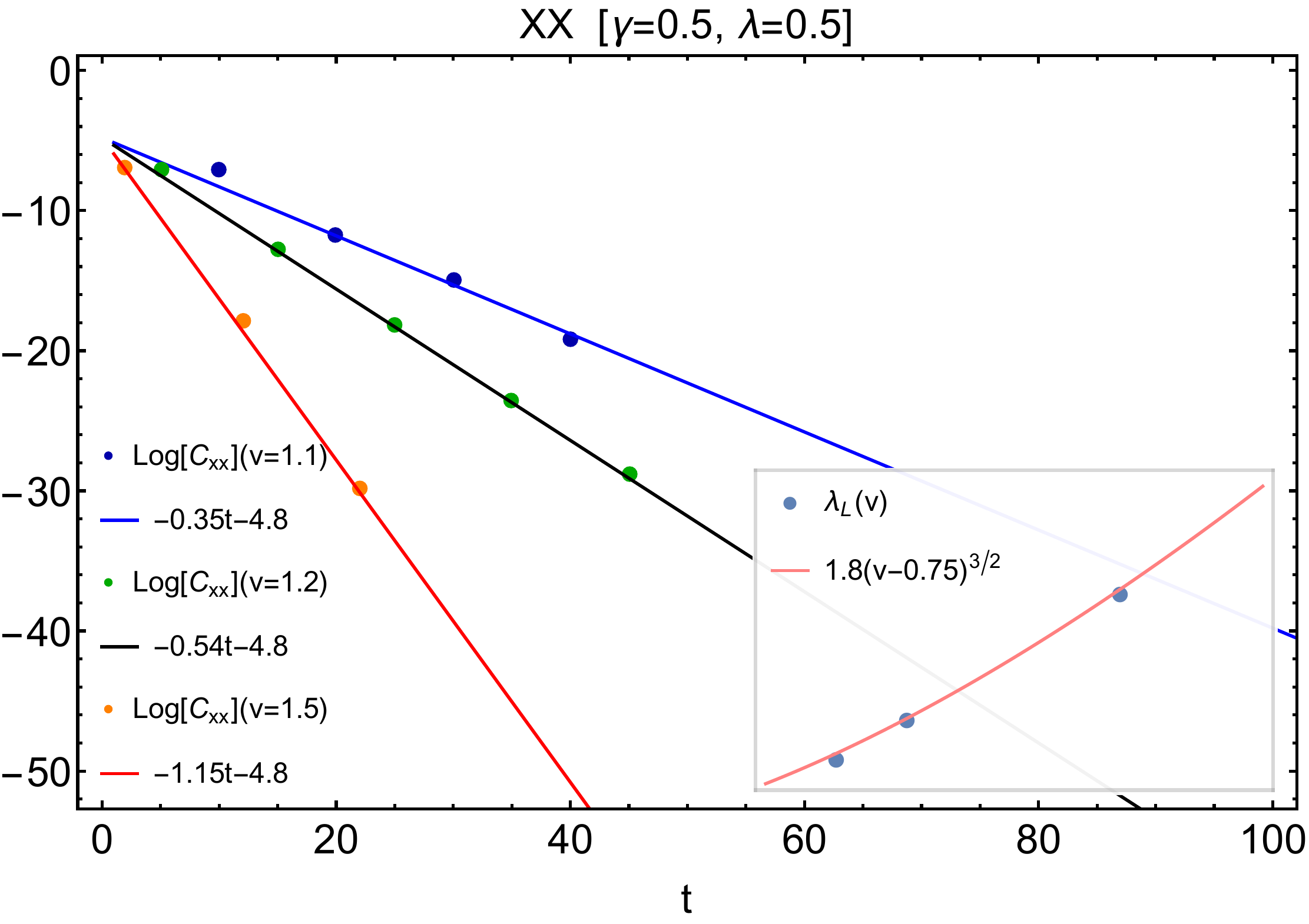}
	\includegraphics[width=0.4\textwidth]{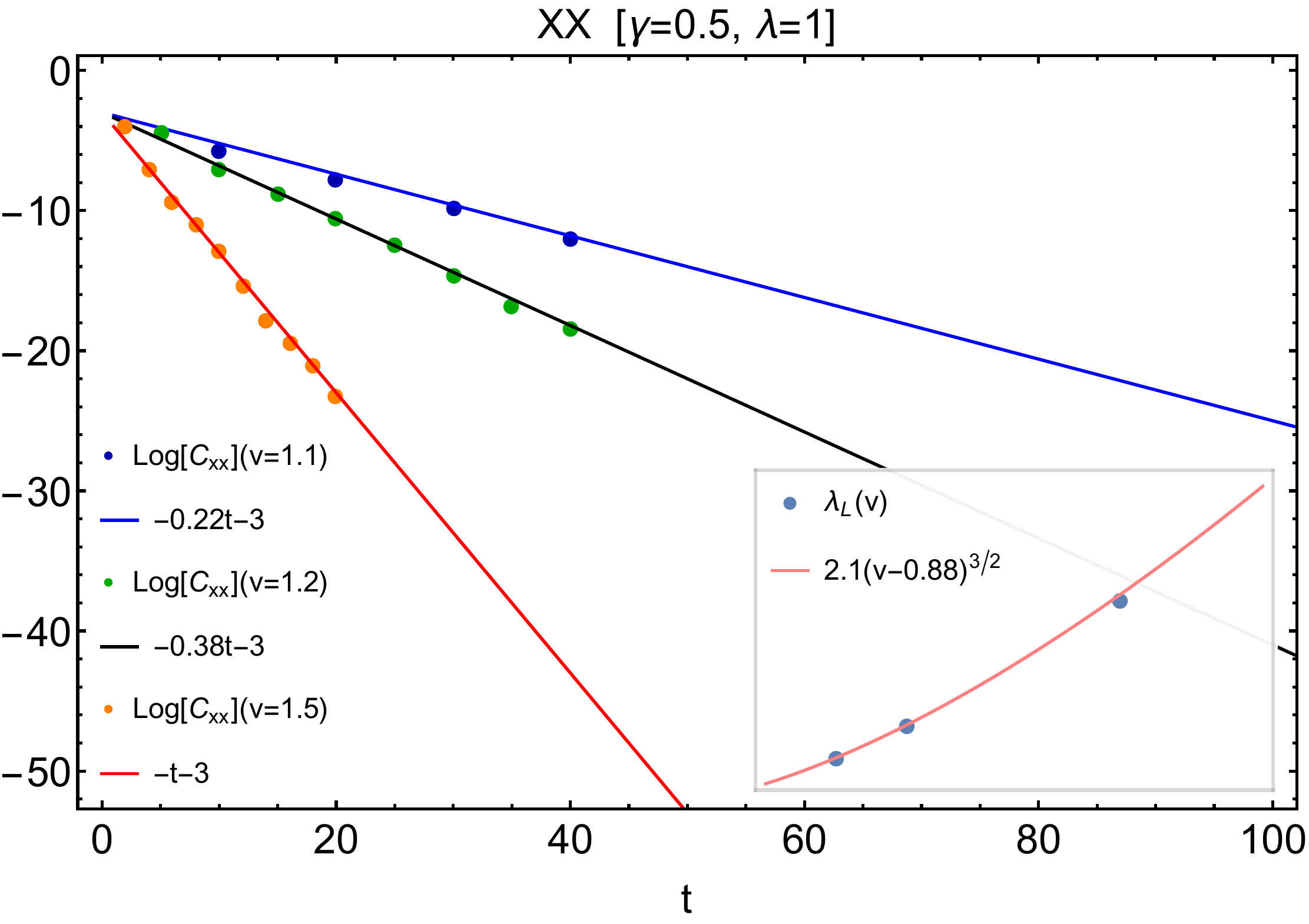}
	\includegraphics[width=0.4\textwidth]{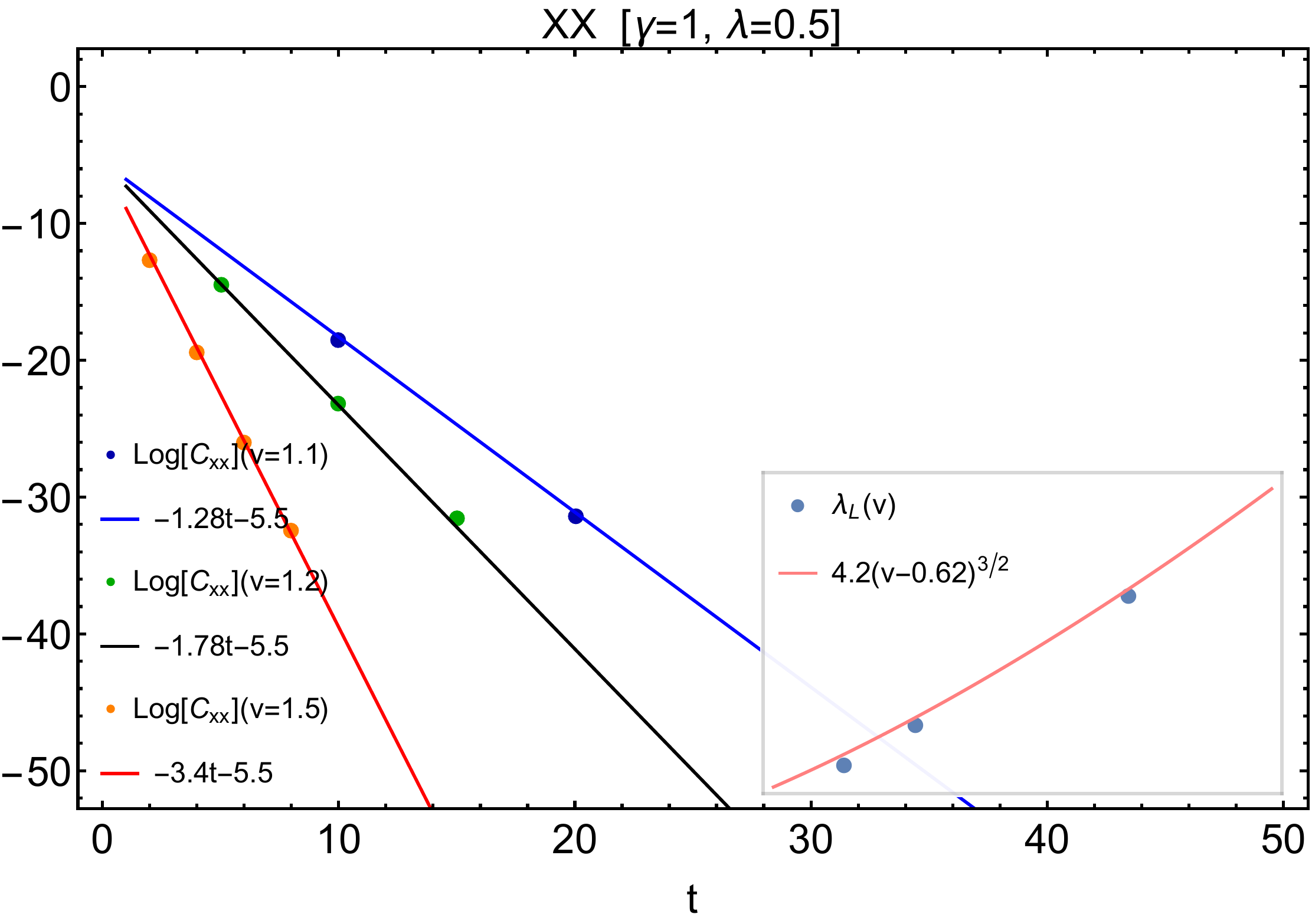}
	\caption{Fitting results of the universal form and numerical data of $C_{xx}$. Here we pick three models with different $v_B$, and all other models and OTOC with nonlocal operators have been verified either. The dots outside are numerical data by calculating the OTOC along velocity-fixed rays with $v=1.1, 1.2, 1.5$ respectively. And the solid lines are fitting forms of $-at+b$, $a$ is the velocity-dependent Lyapunov exponents $\lambda_L(v)$ we need to extract. The inset shows how this three sets of extracted data fitted with $\sim(v-v_B)^{1+p}$ as a function of $v$. We can see the numerical data fits quite well.}\label{fig:NonLWF}
\end{figure}

Having known that the butterfly velocity of different operators is not changed for same model, we can continue to check whether the universal form about the wavefront behavior still holds in nonlocal case. Here we only show the results $C_{xx}$ for same models as local cases due to the limit of space, but we have confirmed that all nonlocal $C(t)$ support the conjecture quite well. The results are illustrated in Fig.\ref{fig:NonLWF}. And we can see that the fitting results of numerical data are quite well in the insets, since their butterfly velocity $v_B$ can be obtained in Fig.\ref{fig:xxsp}. Therefore, we have checked this form with all kinds of OTOC in all phases of XY model, and the results are all positive.

Furthermore, the time evolution of OTOC with nonlocal operators can also be analyzed, including their early time and long time power law behaviors. First, about the early time part, we exhibit the detailed plots in Appendix~\ref{app:plots}, and summarize the results in Table.\ref{tab:et}. Note that since $l=1$ is not special any more in nonlocal operator case, we only plot results of $l=2, 3 ,4$ for clearer vision. Here the sign of OTOC is not a problem because $C(t)$ are relatively close to 0 at early time, so the sign of $|F(t)|$ must be positive. These results are actually not beyond our expectation because all of them agree with the HBC formula. Thus, we don't need to care much about this part.

\begin{figure}[!htbp]
	\centering
	\includegraphics[width=0.45\textwidth]{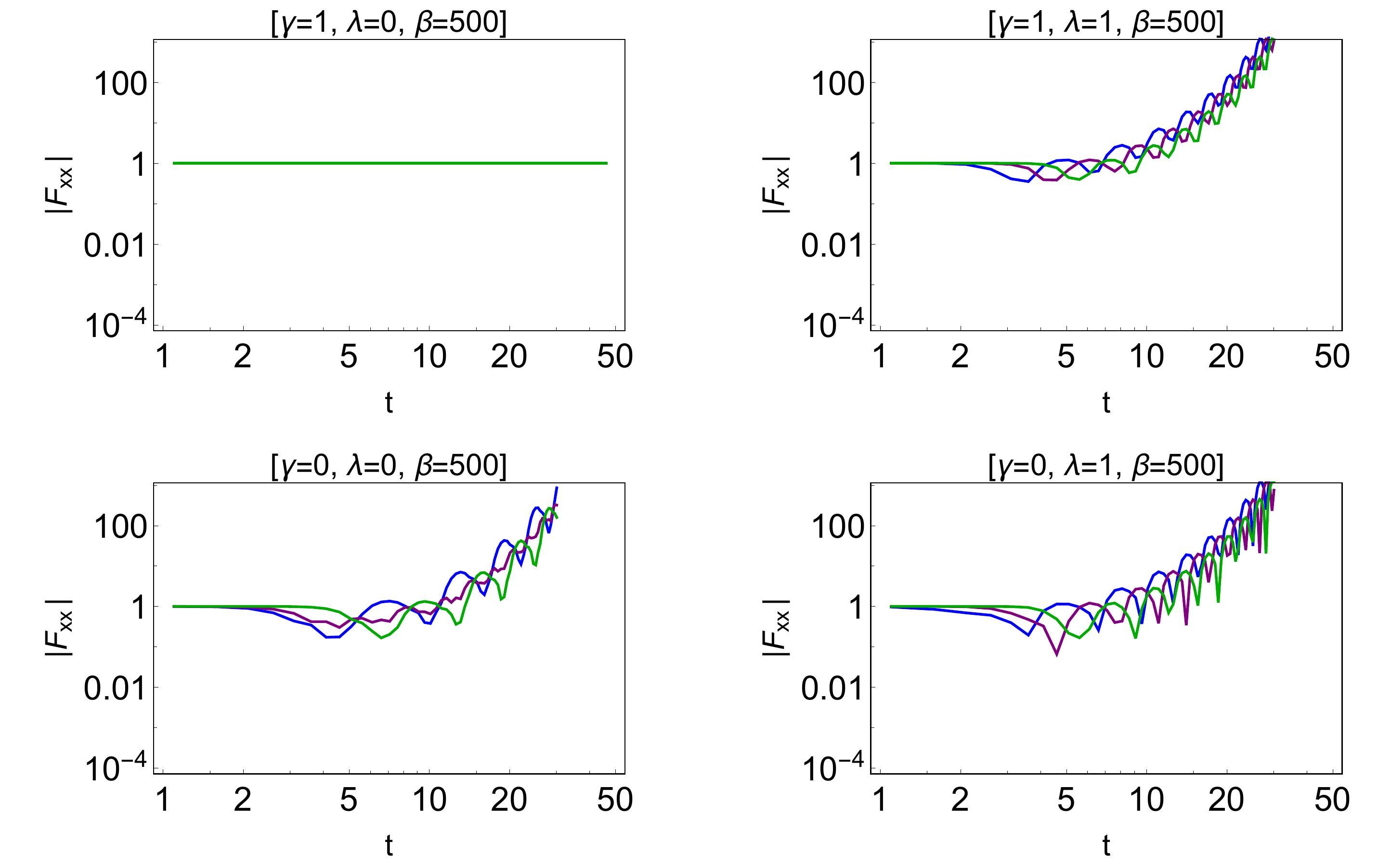}
	\caption{Late time behavior of $|F_{xx}|$ at low temperature $\beta=500$. Blue, purple, green lines correspond to fixed locations $l=2, 3, 4$ respectively. OTOC is observed to be divergent at late time.}\label{fig:500}
\end{figure}

However, the long time behavior is somehow more subtle. In \cite{Lin:2018tce}, the author found that for quantum Ising chain at critical point, $|F_{xx}(t)|$ exhibits nontrivial $t^{-1/4}$ decay at long time, here we confirm this result, and study the long time power law behaviors in other regions. The results are shown in Appendix~\ref{app:plots} and Table.\ref{tab:lt}. Surprisingly, they show quite different power law behaviors with different choices of $\gamma$ and $\lambda$. We know that at late time, $|F(t)|$ with nonlocal operators approach 0, which means $C(t)$ is getting closer to the saturation value 1. Thus, the power law behaviors indicate that these different operators show different rate of saturation. Especially $|F_{xz}(t)|$ and $|F_{yz}(t)|$, which describe how nonlocal operators and local operators interact with each other, show no decay for quantum Ising chain at critical point. Thus, this observation indicates that for these two kinds of OTOC, $C(t)$ will be constant at very long time, not grow at all. Moreover, our results show that they are independent of location $l$ when $\beta=0$, but if we set $\beta$ to a bigger value, the pattern of OTOC is somehow quite omplex, some results of $|F_{xx}|$ are shown in Fig.\ref{fig:500}, we can see it's divergent at late time. The reason is not clear for now, since we use numerical method to do the calculation but not an analytic form. Therefore we don't get a universal description of long time behaviors of OTOC with nonlocal operators, but we do hope our calculation can be helpful to find the final form.

\renewcommand\arraystretch{1.35}
\begin{table}[!htbp]
\centering
\caption{Summary of early time power law growth of OTOC with both local and nonlocal operators in XY model.}
\begin{tabular}{p{1.2cm}<{\centering} p{1.2cm}<{\centering} p{1.2cm}<{\centering} p{1.2cm}<{\centering} p{1.2cm}<{\centering} }
\hline\hline
&(1, 1)&(0, 1)&(1, 0)&(0, 0)\\
\cline{1-5}
$C_{xx}$&$t^{4l+2}$&$t^{2l+1+(-1)^l}$&-&$t^{2l+1+(-1)^l}$\\		
$C_{xy}$&$t^{4l}$&$t^{2l+1-(-1)^l}$&-&$t^{2l+1-(-1)^l}$\\
$C_{yy}$&$t^{4l-2}$&$t^{2l+1+(-1)^l}$&-&$t^{2l+1+(-1)^l}$\\		
$C_{xz}$&$t^{4l}$&$t^{2l}$&-&$t^{2l}$\\
$C_{yz}$&$t^{4l-2}$&$t^{2l}$&-&$t^{2l}$\\
$C_{zz}$&$t^{4l-2}$&$t^{2l}$&$t^0$($l$=$1$)&$t^{2l}$\\
\hline
\end{tabular}
\label{tab:et}
\end{table}

\begin{table}[!htbp]
\centering
\caption{Summary of long time power law growth of OTOC with both local and nonlocal operators in XY model.}
\begin{tabular}{p{1.2cm}<{\centering} p{1.2cm}<{\centering} p{1.2cm}<{\centering} p{1.2cm}<{\centering} p{1.2cm}<{\centering} }
\hline\hline
&(1, 1)&(0, 1)&(1, 0)&(0, 0)\\
\cline{1-5}
$|F_{xx}|$&$t^{-1/4}$&$t^{-1/2}$&-&$t^{-1/2}$\\		
$|F_{xy}|$&$t^{-2}$&$t^{-3/2}$&-&$t^{-1/2}$\\
$|F_{yy}|$&$t^{-3}$&$t^{-1/2}$&-&$t^{-1/2}$\\		
$|F_{xz}|$&$t^{0}$&$t^{-2}$&-&$t^{-2}$\\
$|F_{yz}|$&$t^{0}$&$t^{-2}$&-&$t^{-2}$\\
$C_{zz}$&$t^{-1}$&$t^{-1}$&$t^0$($l$=$1$)&$t^{-1}$\\
\hline
\end{tabular}
\label{tab:lt}
\end{table}

\section{Discussion and conclusion} \label{IV}

The study about OTOC in integrable systems is relatively a new idea, and it may reveal plentiful interesting information about how operators evolve in such systems and how the scrambling happens. In this work we mainly focus on the behaviors of OTOC in XY model, including its early time, long time, wavefront parts, together with the check of the conjectured universal form Eq.\eqref{eq:wavefront}. By careful calculation and analysis we find some interesting points about OTOC in this system. First, we observed that the butterfly velocity in XY model is dependent of its anisotropy parameter $\gamma$ and magnetic fiend $\lambda$, but independent of the locality of operators in OTOC. And based on this observation, we proved that for all kinds of OTOC with all choices of parameters in XY model, the conjectured form Eq. \eqref{eq:wavefront} about the wavefront behavior holds. Therefore, it's indeed a viable description of OTOC around wavefront ($v>v_B$) at least for XY model.

Furthermore, we studied about the time and space evolution of OTOC with both local and nonlocal operators in XY model comprehensively. We find some interesting points about their general behavior. (1). When $\gamma=\beta=0$, OTOC with local operators is independent of external magnetic field $\lambda$; (2). For the noncritical point $\gamma=0$, $\lambda=1$, OTOC with local operators will vanish when temperature falls approaching to zero, but it doesn't happen at other typical sets of parameters. In addition to these results, the early time and long time evolution of OTOC with fixed location have been studied either. We find that while early time behavior totally agree with the power law results from HBC formula, the long time behaviors show nontrivial saturation rate for different operators and models, and they are independent of location $l$ when $\beta=0$, but when temperature becomes lower, their behaviors are very complex, which seems not easy to analyze with numerical calculation. In addition, OTOC with both local and nonlocal operators, i.e. $|F_{xz}|$ and $|F_{yz}$ show $t^0$ at long time, which means their long time evolution is constant, instead of approaching 0 as other kinds.

Overall, we have studied many aspects about OTOC in XY model, and provide some evidences to support the conjecture about chaos spreading around wavefront. But more work need to be done in order to better understand the underlying values about these observations and conclusions, which can also be explored in experiments. And analysis of OTOC in more systems are also required to understand more profound nature of chaos in many-body quantum systems.

\begin{acknowledgments}

We thank Peng-Cheng Li for helpful discussion. This work is funded by China Postdoctoral Science Foundation.

\end{acknowledgments}

\appendix

\onecolumngrid

\section{Majorana two-point correlation functions}
\label{app:Majorana}

Following the definition of Majorana representation, we have
\begin{eqnarray}
\langle A_m(t)A_n\rangle=\langle (c_m^{\dagger}(t)+c_m(t))(c_n^{\dagger}+c_n)\rangle\,,\\
\langle A_m(t)B_n\rangle=\langle (c_m^{\dagger}(t)+c_m(t))(c_n^{\dagger}-c_n)\rangle\,,\\
\langle B_m(t)A_n\rangle=\langle (c_m^{\dagger}(t)-c_m(t))(c_n^{\dagger}+c_n)\rangle\,,\\
\langle B_m(t)B_n\rangle=\langle (c_m^{\dagger}(t)-c_m(t))(c_n^{\dagger}-c_n)\rangle\,,
\end{eqnarray}
then use Fourier transformation $c_j=\frac{e^{i\pi/4}}{\sqrt{N}}\sum_{k}e^{ijk}c_k$ and Bogoliubov transformation $c_k={\rm cos}\theta_k \gamma_k +{\rm sin}\theta_k \gamma_{-k}^{\dagger}$, it's straightforward to expand the above forms in momentum space. Finally we can get

\begin{eqnarray}
\langle A_m(t)A_n\rangle&=&\frac{1}{N}\sum_{k}\Big{[}{\rm cos}(\epsilon_kt)-i{\rm sin}(\epsilon_kt){\rm tanh}\frac{\beta\epsilon_k}{2}\Big{]}e^{i(m-n)k}\,,\\
\langle A_m(t)B_n\rangle&=&\frac{1}{N}\sum_{k}\Big{[}{\rm cos}(\epsilon_kt){\rm tanh}\frac{\beta\epsilon_k}{2}-i{\rm sin}(\epsilon_kt)\Big{]}e^{2i\theta_k}e^{i(m-n)k}\,,\\
\langle B_m(t)A_n\rangle&=&-\frac{1}{N}\sum_{k}\Big{[}{\rm cos}(\epsilon_kt){\rm tanh}\frac{\beta\epsilon_k}{2}-i{\rm sin}(\epsilon_kt)\Big{]}e^{-2i\theta_k} e^{i(m-n)k}\,,\\
\langle B_m(t)B_n\rangle&=&-\frac{1}{N}\sum_{k}\Big{[}{\rm cos}(\epsilon_kt)-i{\rm sin}(\epsilon_kt){\rm tanh}\frac{\beta\epsilon_k}{2}\Big{]}e^{i(m-n)k}\,.
\end{eqnarray}

\section{Time evolution of OTOC with nonlocal operators}
\label{app:plots}

\begin{figure}[!htbp]
	\centering
	\includegraphics[width=1\textwidth]{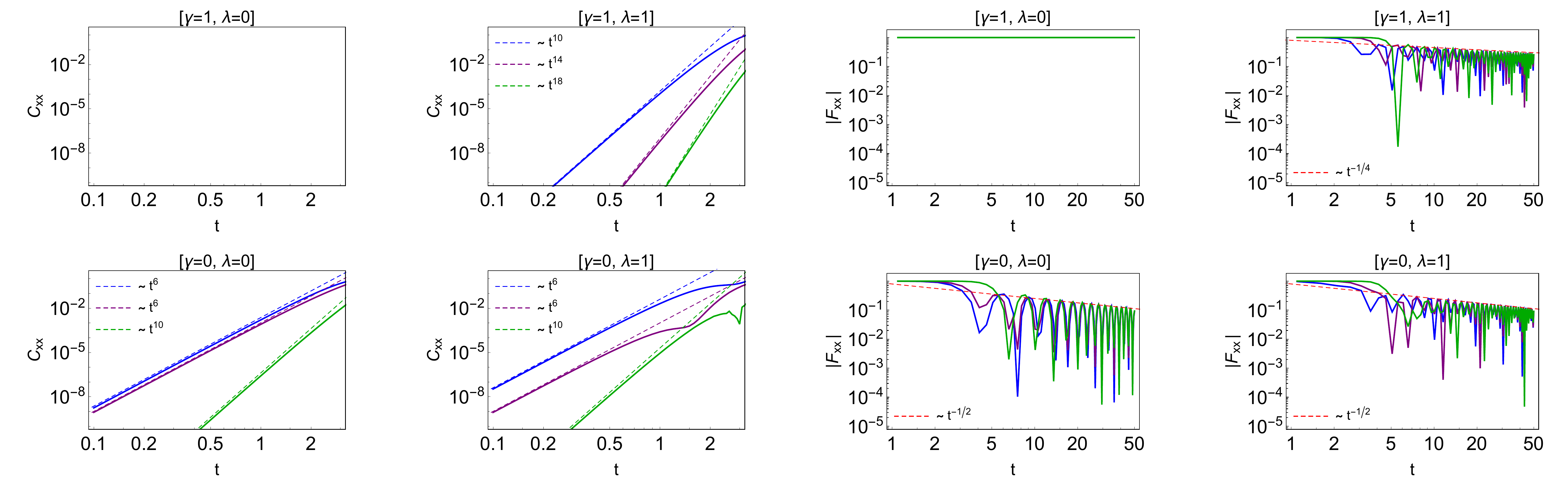}
	\caption{Early and late time evolution of $xx$ OTOC with four sets of parameters and fixed location $l=2, 3, 4$ (blue, purple, green lines).}
\end{figure}

\begin{figure}[!htbp]
	\centering
	\includegraphics[width=1\textwidth]{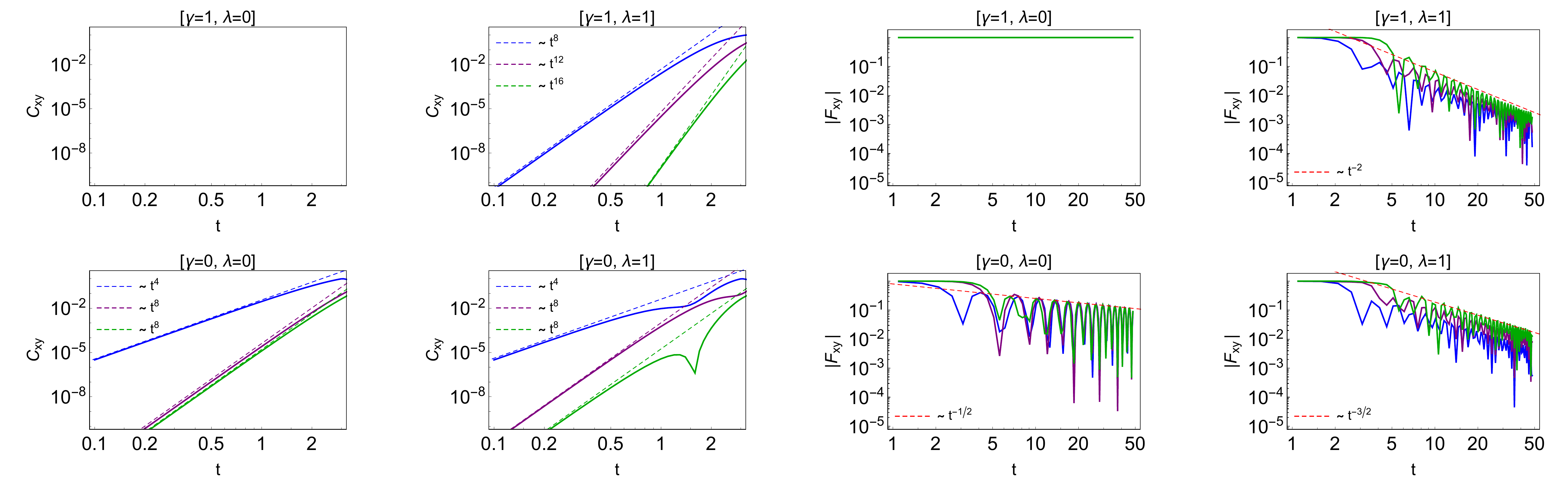}
	\caption{Early and late time evolution of $xy$ OTOC with four sets of parameters and fixed location $l=2, 3, 4$ (blue, purple, green lines).}
\end{figure}

\begin{figure}[!htbp]
	\centering
	\includegraphics[width=1\textwidth]{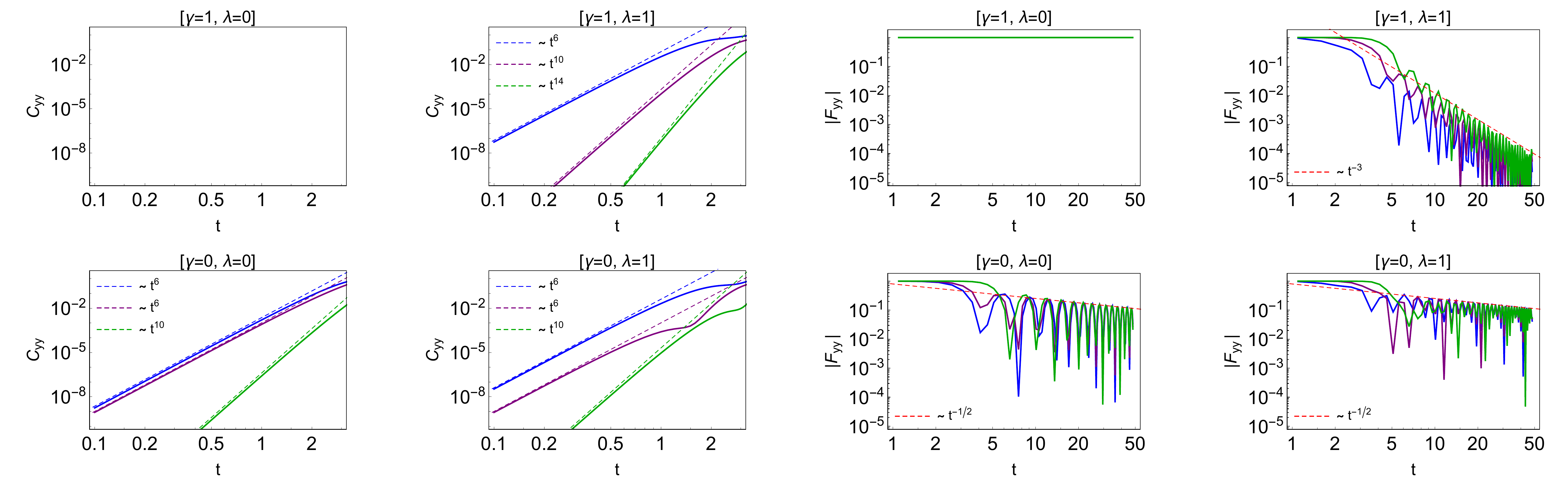}
	\caption{Early and late time evolution of $yy$ OTOC with four sets of parameters and fixed location $l=2, 3, 4$ (blue, purple, green lines).}
\end{figure}

\begin{figure}[!htbp]
	\centering
	\includegraphics[width=1\textwidth]{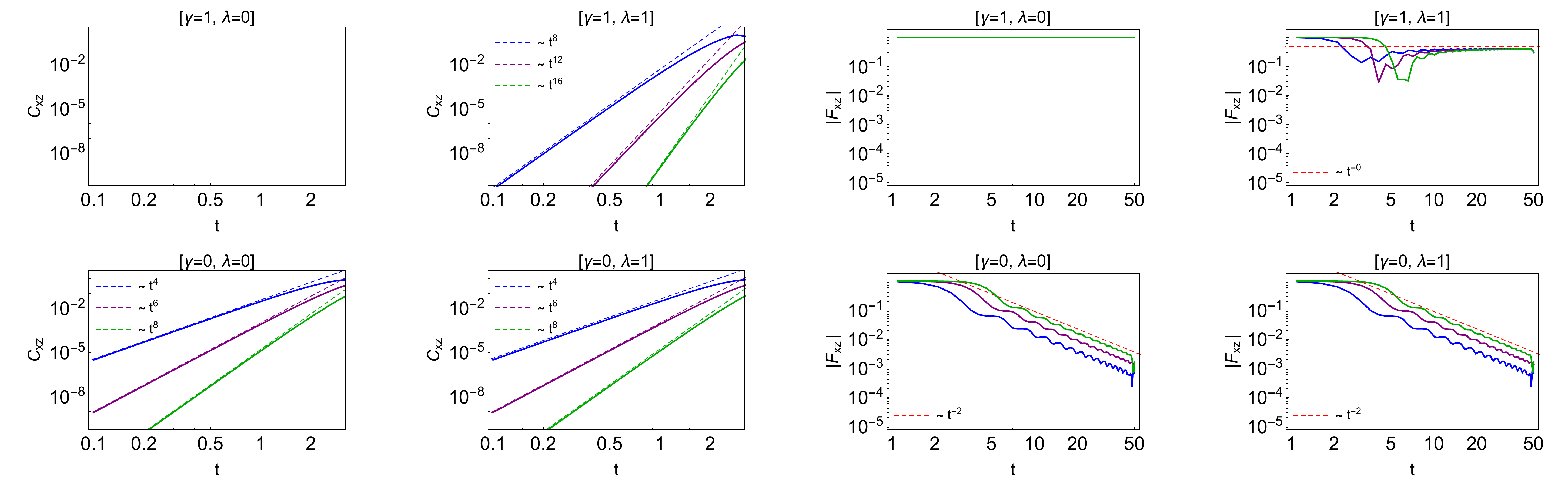}
	\caption{Early and late time evolution of $xz$ OTOC with four sets of parameters and fixed location $l=2, 3, 4$ (blue, purple, green lines).}
\end{figure}

\begin{figure}[!htbp]
	\centering
	\includegraphics[width=1\textwidth]{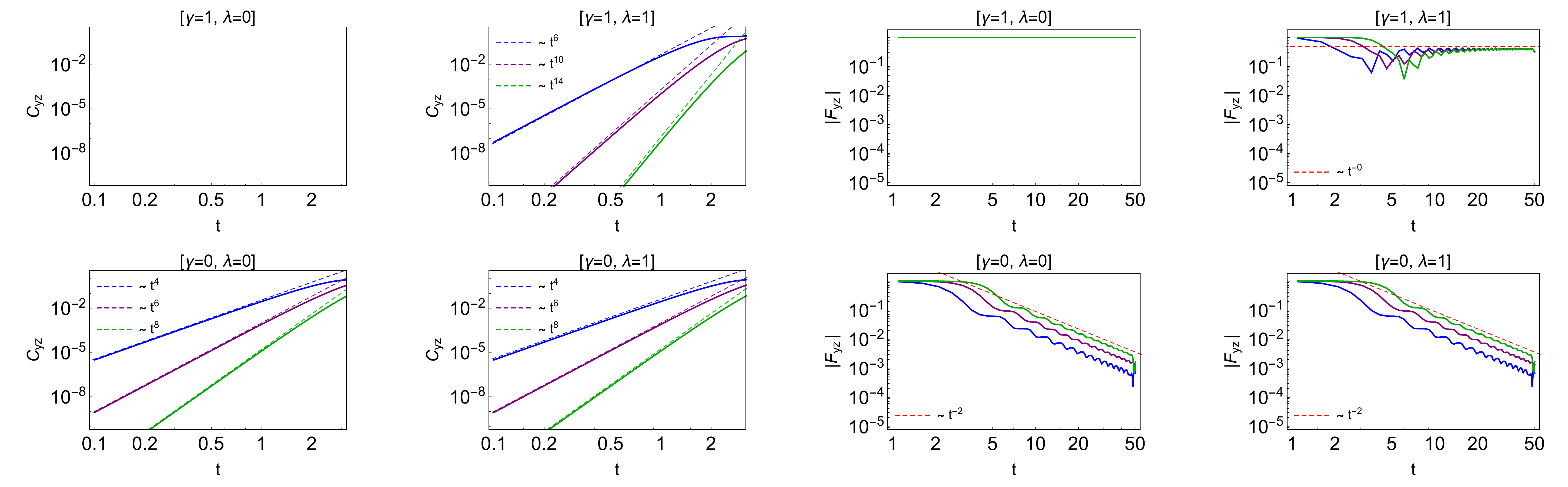}
	\caption{Early and late time evolution of $yz$ OTOC with four sets of parameters and fixed location $l=2, 3, 4$ (blue, purple, green lines).}
\end{figure}

\end{document}